\documentclass[12pt,a4paper,english]{article}
\usepackage[T1]{fontenc}
\usepackage[latin9]{inputenc}
\usepackage{textcomp}
\usepackage{setspace}
\newcommand{\lyxaddress}[1]{
\par {\raggedright #1
\vspace{1.4em}
\noindent\par}
}

\usepackage{babel}

\begin{document}

\title{Analytical Periodic Solutions of Weakly Coupled Map Lattices Using
Recurrence Relations}

\author{Mª Dolores Sotelo Herrera${}^{a}$\&Jesús San Martín${}^{a,b}$}

\date{~}

\maketitle

\lyxaddress{${}^{a}$ Departamento de Matemática Aplicada, E.U.I.T.I., Universidad
Politécnica de Madrid. Ronda de Valencia 3, 28012 Madrid Spain\\
${}^{b}$ Departamento de Física Matemática y de Fluidos, U.N.E.D.
Senda del Rey 9, 28040 Madrid Spain\\
Corresponding author: dsh@dfmf.uned.es}
\begin{abstract}
Analytical periodic solutions for weakly Coupled Map Lattices are
shown in an explicit form as well as in a recurrence relation. The
results establish a link between a matricial representation and recurrence
relations of the solutions.
\end{abstract}

\section{Introduction}

Regular higher order structures, chaotic or not, are very common in
nature. They can be found in physics, chemistry, biology, social sciences
and, in general, in any sytem with interacting elements. Regularity
emerges as a consequence of the interaction among diverse elements
of the system, therefore, to describe them properly, two things should
be considered: the dynamics of every element in the system and the
interactions among them.

A common model used to simulate this situation is Coupled Map Lattices
(CML), dynamical systems with discrete space and time, and continous
state variables. Their origin is in the seminal work of Turing about
morphogenesis \cite{Turing1952}. Later, Kaneko and coworkers studied
them extensively \cite{kaneko1989,kanCluster,kanGCCVL,kanpartt,kaneGCCM}.
CML have proved to be very good in describing collective behaviors
observed in many fields: physics, chemistry, social sciences, biology
\cite{Special1992,Special1997}, that is why they are so extensively
studied. On the other hand a continous non linear system can be studied
by discretizing it \cite{Bakhtier}, what yields a CML. A side effect
of the discretization process is time-computer saving, that plays
an important role when many equations are involved in collective phenomena.
CML have also been used to study the control of spatiotemporal chaos
\cite{Rahmani}.

Mathematically speaking, a one-dimensional CML \cite{librokaneko}
is a $1$-dimensional lattice where each site evolves in time according
to

\[
\begin{array}{c}
X_{i}(n+1)=(1-\alpha)f(X_{i}(n))+\frac{\alpha}{m}\sum_{j=1}^{m}f(X_{j}(n))\\
i=1,...,m\end{array}\]
where $X_{i}(n)$ denotes the state variable for the site $i$ at
time $n$. The parameter $\alpha$ weights the coupling among sites.
Here, periodic boundary conditions are asumed\[
X_{i}(n)=X_{i+m}(n)\;\;\forall i\]

For weak coupling, we write\begin{equation}
\begin{array}{c}
X_{i}(n+1)=(1-\varepsilon\alpha)f(X_{i}(n))+\frac{\alpha\varepsilon}{p}\sum_{j=1}^{p}f(X_{j}(n))\\
i=1,\cdots,p\quad\varepsilon\ll1\end{array}\label{eq:unoo}\end{equation}
where $\alpha$ is the control parameter.

In spite of the big amount of results in CML very few of them are
analytical \cite{Rangarajan}.\textbf{ }Therefore, patterns, periodic
solutions, topological entropy or any other number that measures the
complexity of a dynamical system can not be explicitly deduced. The
goal of this paper is to find the analytical solution of\textbf{ }(\ref{eq:unoo}),
using perturbative methods and recurrence relations. The solution
will be seeked in the form $X_{i}(n)=x_{i}^{*}+\varepsilon A_{i}$
where $x_{i}^{*}\quad i=1,\cdots,p$ are fixed points of $f^{p}$
and $A_{i}$ is a perturbation term that we will explicitly calculate.

\section{Theoretical results}

In order to find the solution of (\ref{eq:unoo}) we will use perturbative
methods, because the system (\ref{eq:unoo}) is weakly coupled. So,
we will seek a solution in the form $X_{i}(n)=x_{i}^{*}+\varepsilon A_{i}$
where $f^{p}(x_{i}^{*})=x_{i}^{*}\quad i=1,\cdots,p$.

If we can find an expresion allowing us to know how the oscilator
$X_{i}(n)$ is turned into $X_{i}(n+q)$ after $q$-iterations, then
we will take advantage of the periodicity of the solution to get $A_{i}$.
We will firstly obtain this expression to reach our goal.

Substituting $x_{i}^{*}+\varepsilon A_{i}$ into (\ref{eq:unoo})
gives \begin{equation}
X_{i}(n+1)=(1-\varepsilon\alpha)f(x_{i}^{*}+\varepsilon A_{i})+\frac{\alpha\varepsilon}{p}\sum_{j=1}^{p}f(x_{j}^{*}+\varepsilon A_{j})\label{eq:ce}\end{equation}
and expanding in powers of $\varepsilon$ (\ref{eq:ce}) becomes

\[
X_{i}(n+1)=f(x_{i})+\varepsilon\left[A_{i}f^{\prime}(x_{i}^{*})-\alpha f(x_{i}^{*})+\frac{\alpha}{p}\sum_{j=1}^{p}f(x_{j}^{*})\right]+O(\varepsilon^{2})\]
or\begin{equation}
\begin{array}{c}
X_{i}(n+1)=f(x_{i}^{*})+\varepsilon t_{i_{1}}+O(\varepsilon^{2})\\
t_{i_{1}}=A_{i}f^{\prime}(x_{i}^{*})-\alpha f(x_{i}^{*})+\frac{\alpha}{p}\sum_{j=1}^{p}f(x_{j}^{*})\end{array}\label{eq:doos}\end{equation}

By using (\ref{eq:doos}) we obtain:

\[
\begin{array}{rl}
X_{i}(n+2)= & (1-\varepsilon\alpha)f(X_{i}(n+1))+\frac{\alpha\varepsilon}{p}\sum_{j=1}^{p}f(X_{j}(n+1))\\
= & (1-\varepsilon\alpha)f\left(f(x_{i}^{*})+\varepsilon t_{i_{1}}\right)+\frac{\alpha\varepsilon}{p}\sum_{j=1}^{p}f(f(x_{j}^{*}))+O(\varepsilon^{2})\\
= & f^{2}(x_{i}^{*})+\varepsilon t_{i_{1}}f^{\prime}(f(x_{i}^{*}))-\varepsilon\alpha f^{2}(x_{i}^{*})+\frac{\alpha\varepsilon}{p}\sum_{j=1}^{p}f^{2}(x_{j}^{*})+O(\varepsilon^{2})\\
= & f^{2}(x_{i}^{*})+\varepsilon\left(t_{i_{1}}f^{\prime}(f(x_{i}^{*}))-\alpha f^{2}(x_{i}^{*})+\frac{\alpha}{p}\sum_{j=1}^{p}f^{2}(x_{j}^{*})\right)+O(\varepsilon^{2})\\
= & f^{2}(x_{i}^{*})+\varepsilon t_{i_{2}}+O(\varepsilon^{2})\end{array}\]
That is\begin{equation}
\begin{array}{c}
X_{i}(n+2)=f^{2}(x_{i}^{*})+\varepsilon t_{i_{2}}+O(\varepsilon^{2})\\
t_{i_{2}}=f^{\prime}(f(x_{i}^{*}))t_{i_{1}}-\alpha f^{2}(x_{i}^{*})+\frac{\alpha}{p}\sum_{j=1}^{p}f^{2}(x_{j}^{*})\end{array}\label{eq:trees}\end{equation}

The former result sugests that \[
\begin{array}{c}
X_{i}(n+q)=f^{q}(x_{i}^{*})+\varepsilon t_{i_{q}}+O(\varepsilon^{2})\\
t_{i_{q}}=f^{\prime}(f^{q-1}(x_{i}^{*}))t_{i_{q-1}}-\alpha f^{q}(x_{i}^{*})+\frac{\alpha}{p}\sum_{j=1}^{p}f^{q}(x_{j}^{*})\end{array}\]
This will be proven in the following theorem.

\paragraph*{Theorem 1}

The CML\[
\begin{array}{c}
X_{i}(n+1)=(1-\varepsilon\alpha)f(X_{i}(n))+\frac{\alpha\varepsilon}{p}\sum_{j=1}^{p}f(X_{j}(n))\\
i=1,\cdots,p\quad\varepsilon\ll1\end{array}\]
becomes\begin{equation}
X_{i}(n+q)=f^{q}(x_{i}^{*})+\varepsilon t_{i_{q}}+O(\varepsilon^{2})\label{eq:asub1}\end{equation}
after $\left(q-1\right)$-iterations, where $t_{i_{q}}$ satisfies
the recurrence relation

\begin{equation}
t_{i_{q}}=\left\{ \begin{array}{cc}
A_{i}f^{\prime}(x_{i}^{*})-\alpha f(x_{i}^{*})+\frac{\alpha}{p}\sum_{j=1}^{p}f(x_{j}^{*}) & q=1\\
f^{\prime}(f^{q-1}(x_{i}^{*}))t_{i_{q-1}}-\alpha f^{q}(x_{i}^{*})+\frac{\alpha}{p}\sum_{j=1}^{p}f^{q}(x_{j}^{*}) & q\geq2\end{array}\right.\label{eq:asub2}\end{equation}

\paragraph*{Proof.}

We will prove the theorem by induction.
\begin{enumerate}
\item [i)] By examination of (\ref{eq:doos}) and (\ref{eq:trees}) the
validity of (\ref{eq:asub1}) and (\ref{eq:asub2}) is obtained for
$q=1$ and $q=2$.
\item [ii)] We assume the expression of $t_{i_{p}}$ is valid for $p=q$
and we will prove it is valid for $p=q+1$.

Let us assume the induction hypothesis\[
X_{i}(n+q)=f^{q}(x_{i}^{*})+\varepsilon t_{i_{q}}+O(\varepsilon^{2})\]

Therefore, it results

$\begin{array}{rl}
X_{i}(n+q+1)= & (1-\varepsilon\alpha)f(X_{i}(n+q))+\frac{\alpha\varepsilon}{p}\sum_{j=1}^{p}f(X_{j}(n+q))\\
= & (1-\varepsilon\alpha)f(f^{q}(x_{i}^{*})+\varepsilon t_{i_{q}}+O(\varepsilon^{2}))+\frac{\alpha\varepsilon}{p}\sum_{j=1}^{p}f(f^{q}(x_{j}^{*})+\varepsilon t_{i_{q}}+O(\varepsilon^{2}))\\
= & (1-\varepsilon\alpha)\left(f(f^{q}(x_{i}^{*}))+f^{\prime}(f^{q}(x_{i}^{*}))\varepsilon t_{i_{q}}\right)+\frac{\alpha\varepsilon}{p}\sum_{j=1}^{p}f(f^{q}(x_{j}^{*}))+O(\varepsilon^{2})\\
= & f^{q+1}(x_{i}^{*})+f^{\prime}(f^{q}(x_{i}^{*}))\varepsilon t_{i_{q}}-\varepsilon\alpha f^{q+1}(x_{i}^{*})+\frac{\alpha\varepsilon}{p}\sum_{j=1}^{p}f^{q+1}(x_{j}^{*})+O(\varepsilon^{2})\\
= & f^{q+1}(x_{i}^{*})+\varepsilon\left(f^{\prime}(f^{q}(x_{i}^{*}))\varepsilon t_{i_{q}}-\alpha f^{q+1}(x_{i}^{*})+\frac{\alpha}{p}\sum_{j=1}^{p}f^{q+1}(x_{j}^{*})\right)+O(\varepsilon^{2})\\
= & f^{q+1}(x_{i}^{*})+\varepsilon t_{i_{q+1}}+O(\varepsilon^{2})\end{array}$

as we wanted to prove\textbf{.}

\end{enumerate}
~

To get a complete solution of the problem we need an explicit (non-recursive)
expression for $t_{i_{q}}$. This is obtained in the following theorem.

\paragraph*{Theorem 2.}

The solution of the recurrence relation\[
t_{i_{q}}=f^{\prime}(f^{q-1}(x_{i}^{*}))t_{i_{q-1}}-\alpha f^{q}(x_{i}^{*})+\frac{\alpha}{p}\sum_{j=1}^{p}f^{q}(x_{j}^{*})\quad q\geq2\]
is given by\begin{equation}
\begin{array}{l}
t_{i_{q}}=A_{i}\left(f^{q}(x_{i}^{*})\right)^{\prime}-\alpha\left(\sum_{n=1}^{q-1}f^{n}(x_{i}^{*})\prod_{k=n}^{q-1}f^{\prime}(f^{k}(x_{i}^{*}))\right)+\\
+\frac{\alpha}{p}\sum_{n=1}^{q-1}\left(\sum_{j=1}^{p}f^{n}(x_{j}^{*})\prod_{k=n}^{q-1}f^{\prime}(f^{k}(x_{i}^{*}))\right)-\alpha f^{q}(x_{i}^{*})+\frac{\alpha}{p}\sum_{l=1}^{p}f^{q}(x_{l}^{*})\end{array}\label{eq:dos}\end{equation}

\paragraph*{Proof.}

We will prove the theorem by induction.
\begin{enumerate}
\item [i)]Let us see that the expression is valid for $q=2$.

By using (\ref{eq:asub2})  results

$t_{i_{2}}=f^{\prime}(f(x_{i}^{*}))t_{i_{1}}-\alpha f^{2}(x_{i}^{*})+\frac{\alpha}{p}\sum_{j=1}^{p}f^{2}(x_{j}^{*})$

$t_{i_{2}}=f^{\prime}(f(x_{i}))\left(A_{i}f^{\prime}(x_{i}^{*})-\alpha f(x_{i}^{*})+\frac{\alpha}{p}\sum_{j=1}^{p}f(x_{j}^{*})\right)-\alpha f^{2}(x_{i}^{*})+\frac{\alpha}{p}\sum_{j=1}^{p}f^{2}(x_{j}^{*})$

\begin{equation}
t_{i_{2}}=A_{i}f^{\prime}(x_{i}^{*})f^{\prime}(f(x_{i}^{*}))-\alpha f(x_{i}^{*})f^{\prime}(f(x_{i}^{*}))+\frac{\alpha}{p}f^{\prime}(f(x_{i}^{*}))\sum_{j=1}^{p}f(x_{j}^{*})-\alpha f^{2}(x_{i}^{*})+\frac{\alpha}{p}\sum_{j=1}^{p}f^{2}(x_{j}^{*})\label{eq:tres}\end{equation}

Subtituting $q=2$ into (\ref{eq:dos}) gives

$t_{i_{2}}=A_{i}\left(f^{2}(x_{i}^{*})\right)^{\prime}-\alpha\left(f(x_{i}^{*})f^{\prime}(f(x_{i}^{*}))\right)+\frac{\alpha}{p}\left(\sum_{j=1}^{p}f(x_{j}^{*})f^{\prime}(f(x_{i}^{*}))\right)-\alpha f^{2}(x_{i}^{*})+\frac{\alpha}{p}\sum_{l=1}^{p}f^{2}(x_{l}^{*})$

This result coincides with the one obtained in (\ref{eq:tres}).

\item [ii)] We assume the induction hypothesis and substituting into

$t_{i_{q+1}}=f^{\prime}(f^{q}(x_{i}^{*}))t_{i_{q}}-\alpha f^{q+1}(x_{i}^{*})+\frac{\alpha}{p}\sum_{j=1}^{p}f^{q+1}(x_{j}^{*})$

results

$\begin{array}{rl}
t_{i_{q+1}}= & f^{\prime}(f^{q}(x_{i}^{*}))\left[A_{i}\left(f^{q}(x_{i}^{*})\right)^{\prime}-\alpha\left(\sum_{n=1}^{q-1}f^{n}(x_{i}^{*})\prod_{k=n}^{q-1}f^{\prime}(f^{k}(x_{i}^{*}))\right)\right]\\
+ & f^{\prime}(f^{q}(x_{i}^{*}))\left[\frac{\alpha}{p}\sum_{n=1}^{q-1}\left(\sum_{j=1}^{p}f^{n}(x_{j}^{*})\prod_{k=n}^{q-1}f^{\prime}(f^{k}(x_{i}^{*}))\right)-\alpha f^{q}(x_{i}^{*})+\frac{\alpha}{p}\sum_{l=1}^{p}f^{q}(x_{l}^{*})\right]\\
- & \alpha f^{q+1}(x_{i}^{*})+\frac{\alpha}{p}\sum_{j=1}^{p}f^{q+1}(x_{j}^{*})\\
= & A_{i}f^{\prime}(f^{q}(x_{i}^{*}))\left(f^{q}(x_{i}^{*})\right)^{\prime}-\alpha\left(f^{\prime}(f^{q}(x_{i}^{*}))\sum_{n=1}^{q-1}f^{n}(x_{i}^{*})\prod_{k=n}^{q-1}f^{\prime}(f^{k}(x_{i}^{*}))\right)\\
+ & \frac{\alpha}{p}\sum_{n=1}^{q-1}f^{\prime}(f^{q}(x_{i}^{*}))\left(\sum_{j=1}^{p}f^{n}(x_{j}^{*})\prod_{k=n}^{q-1}f^{\prime}(f^{k}(x_{i}^{*}))\right)-\alpha f^{\prime}(f^{q}(x_{i}^{*}))f^{q}(x_{i}^{*})\\
+ & \frac{\alpha}{p}f^{\prime}(f^{q}(x_{i}^{*}))\sum_{l=1}^{p}f^{q}(x_{l}^{*})-\alpha f^{q+1}(x_{i}^{*})+\frac{\alpha}{p}\sum_{j=1}^{p}f^{q+1}(x_{j}^{*})\\
= & A_{i}\left(f^{q+1}(x_{i}^{*})\right)^{\prime}-\alpha\left(\sum_{n=1}^{q-1}f^{n}(x_{i}^{*})\prod_{k=n}^{q}f^{\prime}(f^{k}(x_{i}^{*}))\right)\\
+ & \frac{\alpha}{p}\sum_{n=1}^{q-1}\left(\sum_{j=1}^{p}f^{n}(x_{j}^{*})\prod_{k=n}^{q}f'^{\prime}(f^{k}(x_{i}^{*}))\right)-\alpha f^{\prime}(f^{q}(x_{i}^{*}))f^{q}(x_{i}^{*})\\
+ & \frac{\alpha}{p}f^{\prime}(f^{q}(x_{i}^{*}))\sum_{l=1}^{p}f^{q}(x_{l}^{*})-\alpha f^{q+1}(x_{i})+\frac{\alpha}{p}\sum_{j=1}^{p}f^{q+1}(x_{j})\\
= & A_{i}\left(f^{q+1}(x_{i}^{*})\right)^{\prime}-\alpha\left(\sum_{n=1}^{q}f^{n}(x_{i}^{*})\prod_{k=n}^{q}f^{\prime}(f^{k}(x_{i}^{*}))\right)\\
+ & \frac{\alpha}{p}\sum_{n=1}^{q}\left(\sum_{j=1}^{p}f^{n}(x_{j}^{*})\prod_{k=n}^{q}f^{\prime}(f^{k}(x_{i}^{*}))\right)-\alpha f^{q+1}(x_{i}^{*})+\frac{\alpha}{p}\sum_{j=1}^{p}f^{q+1}(x_{j}^{*})\end{array}$

\end{enumerate}
~

As we have previously indicated, now we can obtain $A_{i}$ by using
theorems 1 and 2 and the periodic boundary conditions.

\paragraph*{Theorem 3.}

Let $\left\{ x_{1}^{*},x_{2}^{*},...,x_{p}^{*}\right\} $ be a $p$-periodic
orbit of a $C^{2}$ function $f$ such that $f^{p^{\prime}}(x_{i}^{*})\neq1\quad i=1,\cdots,p$
then the CML given by \begin{equation}
\begin{array}{rl}
X_{i}(n+1) & =(1-\varepsilon\alpha)f(X_{i}(n))+\frac{\alpha\varepsilon}{p}\sum_{j=1}^{p}f(X_{j}(n))\\
i= & 1,\cdots,p\quad\varepsilon\ll1\end{array}\label{eq:cinco}\end{equation}
has a $p$-periodic solution, given by \[
X_{i}(n+j)=x_{i+j}^{*}+\varepsilon A_{i+j}\]
\[
\begin{array}{c}
i=1,\cdots,p\\
j=0,\cdots,p-1\end{array}\]
where\[
A_{i}=\frac{\alpha}{1-\left(f^{p}(x_{i}^{*})\right)^{\prime}}\left[\sum_{n=1}^{p-1}\left(\left(-x_{i+n}^{*}+\frac{1}{p}\sum_{j=1}^{p}x_{j}^{*}\right)\prod_{k=n}^{p-1}f^{\prime}(x_{i+k}^{*})\right)+\left(-x_{i}^{*}+\frac{1}{p}\sum_{l=1}^{p}x_{l}^{*}\right)\right]\]
\[
i=1,\dots,p\]
with periodic boundary conditions\[
\begin{array}{c}
A_{i+p}=A_{i}\\
x_{i+p}^{*}=x_{i}^{*}\end{array}\quad\forall i\]

\paragraph*{Proof}

$X_{i}(n)=x_{i}^{*}+\varepsilon A_{i}$ is $p$-periodic orbit when
$X_{i}(n+p)=x_{i}^{*}+\varepsilon A_{i}$.

By using theorem 2 it turns out that

\[
\begin{array}{rl}
X_{i}(n+p)= & f^{p}(x_{i}^{*})+\varepsilon\left[A_{i}\left(f^{p}(x_{i}^{*})\right)^{\prime}\right.\\
 & -\alpha\left(\sum_{n=1}^{p-1}f^{n}(x_{i}^{*})\prod_{k=n}^{p-1}f^{\prime}(f^{k}(x_{i}^{*}))\right)\\
 & +\frac{\alpha}{p}\sum_{n=1}^{p-1}\left(\sum_{j=1}^{p}f^{n}(x_{j}^{*})\prod_{k=n}^{p-1}f^{\prime}(f^{k}(x_{i}^{*}))\right)\\
 & \left.-\alpha f^{p}(x_{i}^{*})+\frac{\alpha}{p}\sum_{l=1}^{p}f^{p}(x*_{l})\right]+O(\varepsilon^{2})\end{array}\]
As \[
\left\{ \begin{array}{c}
f^{p}(x_{i}^{*})=x_{i}^{*}\\
f^{n}(x_{i}^{*})=x_{i+n}^{*}\\
\sum_{j=1}^{p}f^{n}(x_{j}^{*})=\sum_{j=1}^{p}x_{j}^{*}\end{array}\right.\]
the following equality is obtained:

\[
\begin{array}{rl}
x_{i}^{*}+\varepsilon A_{i}= & x_{i}^{*}+\varepsilon\left[A_{i}\left(f^{p}(x_{i}^{*})\right)^{\prime}\right.\\
 & -\alpha\left(\sum_{n=1}^{p-1}x_{i+n}^{*}\prod_{k=n}^{p-1}f^{\prime}(f^{k}(x_{i}^{*}))\right)\\
 & +\frac{\alpha}{p}\sum_{n=1}^{p-1}\left(\sum_{j=1}^{p}x_{j}^{*}\prod_{k=n}^{p-1}f^{\prime}(f^{k}(x_{i}^{*}))\right)\\
 & \left.-\alpha x_{i}^{*}+\frac{\alpha}{p}\sum_{l=1}^{p}x_{l}^{*}\right]+O(\varepsilon^{2})\end{array}\]
Eliminating $x_{i}^{*}$ and separating terms of $A_{i}$ yields

\[
\begin{array}{rl}
\left[1-\left(f^{p}(x_{i}^{*})\right)^{\prime}\right]A_{i}= & -\alpha\sum_{n=1}^{p-1}x_{i+n}^{*}\prod_{k=n}^{p-1}f^{\prime}(f^{k}(x_{i}^{*}))\\
 & +\frac{\alpha}{p}\sum_{n=1}^{p-1}\left(\sum_{j=1}^{p}x_{j}^{*})\prod_{k=n}^{p-1}f^{\prime}(x_{i+k}^{*})\right)\\
 & -\alpha x_{i}^{*}+\frac{\alpha}{p}\sum_{l=1}^{p}x_{l}^{*}\\
\\= & \sum_{n=1}^{p-1}\left[\left(-\alpha x_{i+n}^{*}+\frac{\alpha}{p}\sum_{j=1}^{p}x_{j}^{*}\right)\prod_{k=n}^{p-1}f^{\prime}(x_{i+k}^{*})\right]\\
 & -\alpha x_{i}^{*}+\frac{\alpha}{p}\sum_{l=1}^{p}x_{l}^{*}\end{array}\]
and it is finally obtained that

\begin{equation}
A_{i}=\frac{\alpha}{1-\left(f^{p}(x_{i}^{*})\right)^{\prime}}\left[\sum_{n=1}^{p-1}\left(\left(-x_{i+n}^{*}+\frac{1}{p}\sum_{j=1}^{p}x_{j}^{*}\right)\prod_{k=n}^{p-1}f^{\prime}(x_{i+k}^{*})\right)+\left(-x_{i}^{*}+\frac{1}{p}\sum_{l=1}^{p}x_{l}^{*}\right)\right]\label{eq:final}\end{equation}

The result is valid while $1-\left(f^{p}(x_{i}^{*})\right)^{\prime}$
is $O(1)$, so that (\ref{eq:doos}), obtained after expanding in
powers of $\varepsilon$, holds. The solution given by theorem 3 represents
a propagating wave in the CML. Traveling waves in CML may explain
ordered structures observed in nature \cite{Qian}.

The expresion (\ref{eq:final}) coincides with the one obtained in
\textbf{\cite{nosotros}} (see theorem 4 therein), thus providing
an alternative way of facing CML. In \cite{nosotros} a matrix representation
was given, meanwhile here the solution is obtained from a recurrence
relation.\textbf{ }Therefore a link between both representation is
established. Two different approaches to the same problem provide
a powerful framework to draw theoretical conclusions, in particular
in areas like CML where most of the published results are numerical.

The analytical approach allows to derive theoretical expressions for
synchronizated states, travelling waves, period doubling cascades,
intermittency, and so on \cite{nosotros}. However, depending on the
problem, recurrence relations or matricial representation may become
a more or less useful tool in their analysis.

\section{Discussion and conclusions}

We have found analytical periodic solutions for weakly Coupled Map
Lattices,\textbf{ }using recurrence relations.\textbf{ }From recurrence
relations an explicit solution is deduced, that coincides with the
one obtained using matricial approach, therefore, the link between
both techniques is established.\textbf{ }The solution is very general
because the result is valid for an arbitrary number of oscillators,
and furthermore the individual dynamics of every oscillator is assumed
to be ruled by an arbitrary $C^{2}$ function.

Analytical reults are particularly important for two reasons:
\begin{description}
\item [{First.-}] Analytical expressions can obviously be managed with
mathematical tools and new physical phenomena can be deduced from
them.
\item [{Second.-}] Spurious results, due to finite precision in computer
simulations, are avoided \cite{Grebogi,Czou2000,CZhou2002}. 
\end{description}
Another couple of interesting facts must be pointed out. On the one
hand, the recurrence relation allows for an exact and efficient numerical
solution in a computer. On the other hand, as the solution is obtained
for an arbitrary number of oscillators, limits tending to infinity
can be calculated. This is an important point in the study of bifurcation
cascades and of the onset the turbulence in fluids, where a finite
number of elements is not adequate to described the phenomena.

\end{document}